\documentclass[twocolumn]{aastex63}

\usepackage{gensymb}
\usepackage{wrapfig}
\usepackage{booktabs}

\usepackage{xcolor}

\shorttitle{CME Deflection and East-West Asymmetry of ESP Intensity in Solar Cycles 23 and 24}
\shortauthors{Santa Fe Dueñas et al.}

\graphicspath{{./}{figures/}}

\begin{document}

\title{CME Deflection and East-West Asymmetry of Energetic Storm Particle Intensity during Solar Cycles 23 and 24}

\correspondingauthor{Adolfo {Santa Fe Dueñas}}
\email{adolfo.santafeduenas@swri.org}

\author[0000-0002-6035-1254]{A. Santa Fe Dueñas}

\affiliation{Space Science and Engineering Division, Southwest Research Institute, San Antonio, TX 78228, USA}

\author[0000-0002-2504-4320]{R. W. Ebert}
\affiliation{Space Science and Engineering Division, Southwest Research Institute, San Antonio, TX 78228, USA}
\affiliation{Department of Physics and Astronomy, University of Texas at San Antonio, San Antonio, TX, USA}

\author[0000-0003-4695-8866]{Gang {Li}}
\affiliation{Department of Space Science and CSPAR, University of Alabama in Huntsville, Huntsville, AL 35805, USA}

\author[0000-0002-9829-3811]{Zheyi {Ding}}
\affiliation{Centre for Mathematical Plasma Astrophysics, KU Leuven, 3001 Leuven, Belgium}

\author[0000-0001-9323-1200]{M. A. Dayeh}
\affiliation{Space Science and Engineering Division, Southwest Research Institute, San Antonio, TX 78228, USA}
\affiliation{Department of Physics and Astronomy, University of Texas at San Antonio, San Antonio, TX, USA}

\author[0000-0002-7318-6008]{M. I. Desai}
\affiliation{Space Science and Engineering Division, Southwest Research Institute, San Antonio, TX 78228, USA}
\affiliation{Department of Physics and Astronomy, University of Texas at San Antonio, San Antonio, TX, USA}

\author[0000-0002-6849-5527]{L. K. Jian}
\affiliation{Heliophysics Science Division, NASA Goddard Space Flight Center, Greenbelt, MD 20771, USA}

\begin{abstract}
We investigate the East-West asymmetry in energetic storm particle (ESP) heavy ion intensities at interplanetary shocks driven by coronal mass ejections (CMEs) during solar cycles (SCs) 23 and 24. We use observations from NASA's ACE and STEREO missions of helium (He), oxygen (O), and iron (Fe) intensities from $\sim$0.13 to 3 MeV/nucleon. We examine the longitudinal distribution of ESP intensities and the correlation of ESP intensities with the near-Sun CME speed and the average transit CME speed for eastern and western events. We observed an East-West asymmetry reversal of ESP heavy ion intensities from SC 23 to 24. We have determined that this change in asymmetry is caused by a shift in the heliolongitude distribution of the CME speed ratio (the ratio of CME near-Sun speed to CME average transit speed) from west to east.

\end{abstract}


\section{Introduction} \label{sec:intro}

\begin{figure}[ht!]
\includegraphics[width=\linewidth]{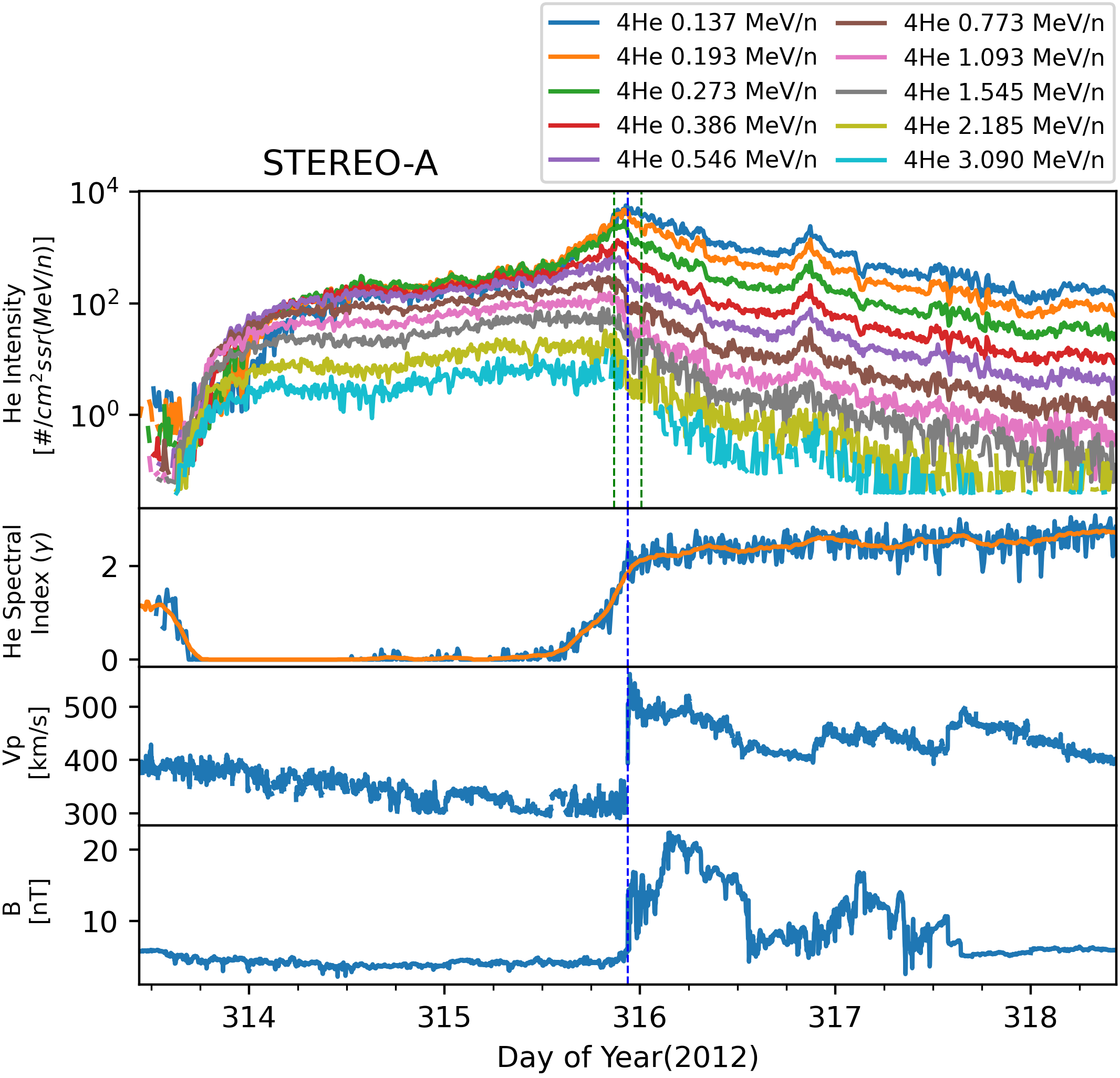} 
\centering
\caption{An example of an ESP event observed by STA at one au on November 10 and 11, 2012. The first panel shows the Helium (He) time-intensity profiles (0.137 - 3.090 MeV/n), the second panel shows the He spectral index from $\sim$ 0.1-3 MeV/nucleon, the solar wind speed is shown in the third panel, and the fourth panel shows the magnetic field strength. The blue vertical lines denote the shock arrival time. The orange line in the spectral index panel is the running average of a 21-sample window. To identify the peak intensity values, a time window is defined by the green vertical lines, which set the limits of the time range to be considered.}
\label{fig:ESP}
\end{figure}

There are two classes of Solar Energetic Particle (SEP) events: impulsive SEP events associated with impulsive X-ray solar flares and gradual SEP events accelerated by Coronal Mass Ejections (CME) driven shocks \citep{Reames2013}. The enhancement of the intensity of energetic particles during the passage of the interplanetary (IP) shock in SEP events is known as an Energetic Storm Particle (ESP) event \citep{Bryant1962}. Figure \ref{fig:ESP} shows an example ESP event observed by the STEREO-A spacecraft. The sudden increase in the solar wind speed and the magnetic field, shown in the third and fourth rows, indicates the arrival of the IP shock, whereas the rise in the Helium (He) intensities at about the shock arrival time is the ESP event.

The properties of ESP events depend on the spacecraft (s/c) position along the IP shock, the shock properties, and the upstream conditions. For instance, \citet{Desai2003} and \citet{Dayeh2009} found a correlation of ESP properties with the upstream seed particle population. \citet{Cane1988} showed how the intensity–time profiles of solar proton events depend on the longitude of the observing s/c relative to the location of solar flare. \citet{Makela2011}, \citet{Reames_2012}, and \citet{Lario2023} found that ESP intensities have a strong correlation with the CME and IP shock speeds. \citet{SantaFe2022} and \citet{Lario2023} showed correlations of the peak intensities of ESP events with the observing s/c heliolongitude and the CME average transit speed. ESP properties are also affected by the strength of the solar cycle \citep[SC,][]{SantaFe2023}. In addition, \citet{SantaFe2022} showed that the ratio of the CME average transit speed to the CME near-Sun speed correlates with the observing s/c heliolongitude. By combining this correlation and the correlation of the ESP peak intensities to the CME average transit speed, the authors obtained an empirical equation that relates the ESP peak intensities and spectral index to the CME near-Sun speed and the observing s/c heliolongitude. However, this heliolongitude distribution of ESP peak intensities shown in \citet{SantaFe2022} is asymmetric, with eastern ESP events having slightly higher intensities than western events. According to \citet{ding2022modeling}, the difference in fluence and peak flux between the East and West during entire Solar Energetic Particle (SEP) events is caused by the time-extended shock acceleration and the geometry of Parker field lines. On the other hand, \citet{ding2023} suggest that the asymmetry in ESP events is due to the injection efficiency of seed populations, which depends on the shock geometry. Both studies propose that Eastern events have higher intensities than Western events, resulting from combining these two effects. In this study, we want to determine if the East-West asymmetry in ESP properties is consistent between SC 23 and the relatively weaker SC 24. Specifically, we compare heavy ion (He, O, and Fe) intensities from ESP events observed during these two SCs and their dependence on the CME near-Sun and average transit speeds, and the parent flare longitude.

\section{Observations} \label{sec:observations}

We use data from instruments on board NASA’s Advanced Composition Explorer \citep[ACE;][]{Stone1998} and Solar Terrestrial Relations Observatory \citep[STEREO;][]{Kaiser2008} missions. ACE is located at the Sun-Earth L1 Lagrange Point. STEREO consists of two spacecraft, each drifting apart from Earth in opposite directions at about 22\degree\ per year, where STEREO-A (STA) orbits ahead of Earth and STEREO-B (STB) behind. Communication with STB was lost in October 2014.

 We surveyed energetic helium (He), oxygen (O), and iron (Fe) ion observations with energies ranging from 0.1 to 3 MeV/nucleon using the Ultra-Low Energy Isotope Spectrometer (ULEIS) on ACE spacecraft during SC 23 ESP events between August 1998 and December 2006. Additionally, we surveyed the same type of observations using Suprathermal Ion Telescope (SIT) on STA and STB and ULEIS on ACE during SC 24 events between January 2009 and December 2019. Several multi-s/c studies have described the inter-calibration process of these instruments \citep[e.g.][]{Muller-Mellin2008,Dresing2009,Cohen2018}. We use the solar wind plasma and magnetic field instruments on STA and STB (PLASTIC; \citet{Galvin2008} and MAG; \citet{Acuna2008}) and ACE (SWEPAM; \citet{McComas1998} and MAG; \citet{Smith1998}) to detect and analyze interplanetary (IP) shocks and their properties. We also examine the solar wind conditions upstream and downstream of these structures. CME speeds near the Sun and parent flare location are obtained from estimates reported in \citet{Gopalswamy_2010} for SC 23 events and from the Space Weather Database of Notifications, Knowledge, and Information (DONKI\footnote{\url{https://kauai.ccmc.gsfc.nasa.gov/DONKI/}}) for SC 24 events, while the CME average transit speeds to 1 au are estimated from the transit time of the event.
\subsection{ESP event selection}
We analyzed the list of events from previous studies \citep{SantaFe2022,SantaFe2023}, where we visually inspected the plots of the events. We selected the events where the energetic particle intensity profiles exhibited a synchronized increase of at least 200\% within $\pm$3 hours of the shock arrival time in the He, O, and Fe intensities at each energy bin (10 bins for He, 11 bins for O, and 11 bins for Fe) between 0.1 - 3 MeV/nucleon. We also ensured no other dominant shocks were present within a 1-day window.

\section{Results}
\label{subsec:Results5}

\begin{figure}[ht!]
\includegraphics[width=\linewidth]{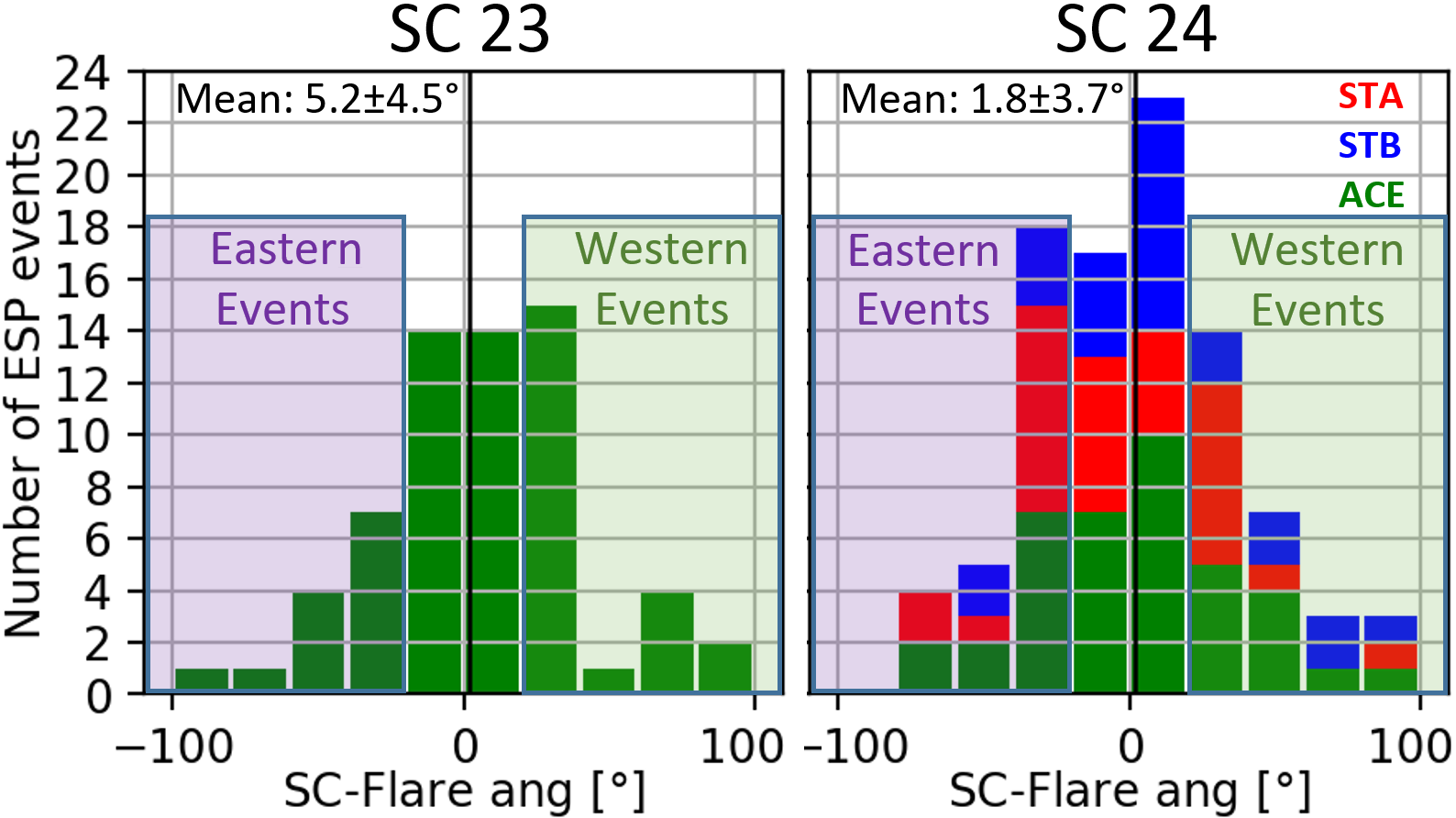} 
\centering
\caption{Helio-longitude distributions of ESP events observed during SC 23 (left) and SC 24 (right). The events are grouped in bins of 20 degrees. The SC 24 ESP events include STA (red), STB (blue), and ACE (green) observations, whereas SC 23 includes only events from ACE. The mean spacecraft-flare angle is shown with a black solid line.}
\label{fig:hist}
\end{figure}

The number of ESP events detected by ACE decreased during the weaker SC 24 compared to the stronger SC 23. However, including ESP events from the STEREO mission, which fortunately began operating during this period, improved the statistics for SC 24. As a result, 76 events from SC 23 and 94 events from SC 24 were selected. The histogram in Figure \ref{fig:hist} shows the distribution of ESP events during SC 23 (left) and SC 24 (right) as a function of the spacecraft-flare angle $\phi$ from \citet{SantaFe2022}. In this context, the CME solar source location is approximated by the solar flare location for each event \citep{Ontiveros_2009}.
The distribution shows a mean helio-longitude of 5.2 $\pm$ 4.5\degree for SC 23 ESP events, whereas for SC 24, the mean helio-longitude is 1.8 $\pm$ 3.7\degree.
Similar results have been shown by \citet{SantaFe2022} and \citet{Lario2023}. 
We have categorized the ESP events into three groups based on their spacecraft-flare angle $\phi$. The Eastern events are those where $\phi \leq -20\degree$, which are shown within the purple shadow region in the figure. The Central events are those where $-20\degree < \phi < 20\degree$. The Western events are those where $\phi \geq 20\degree$, which are shown within the green shadow region in the figure. 

During SC 23, there were 22 Western ESP events and 13 Eastern ESP events. In SC 24, there were 27 Western ESP events and 27 Eastern ESP events.

Using the methodology from \citet{SantaFe2022}, we calculate the average transit speed of each CME event using the formula $\bar V_T=R/(t_{1au}-t_{R_{Sun}})$, where $R$ stands for the distance between the Sun and the observing s/c, $t_{1au}$ is the shock arrival time at the observing s/c (which is about one au away from the Sun), and $t_{R_{Sun}}$ is the CME onset time observed on the Sun by LASCO/C2 at about 1.5 $R_{Sun}$. We calculated the ratio between the average transit speed of CMEs and their near-Sun speeds for ESP events, expressed as $CME\ Speed\ Ratio=\bar V_T/V_i$. We also analyzed its heliolongitudinal distribution. 

\begin{figure}[ht!]
\includegraphics[width=\linewidth]{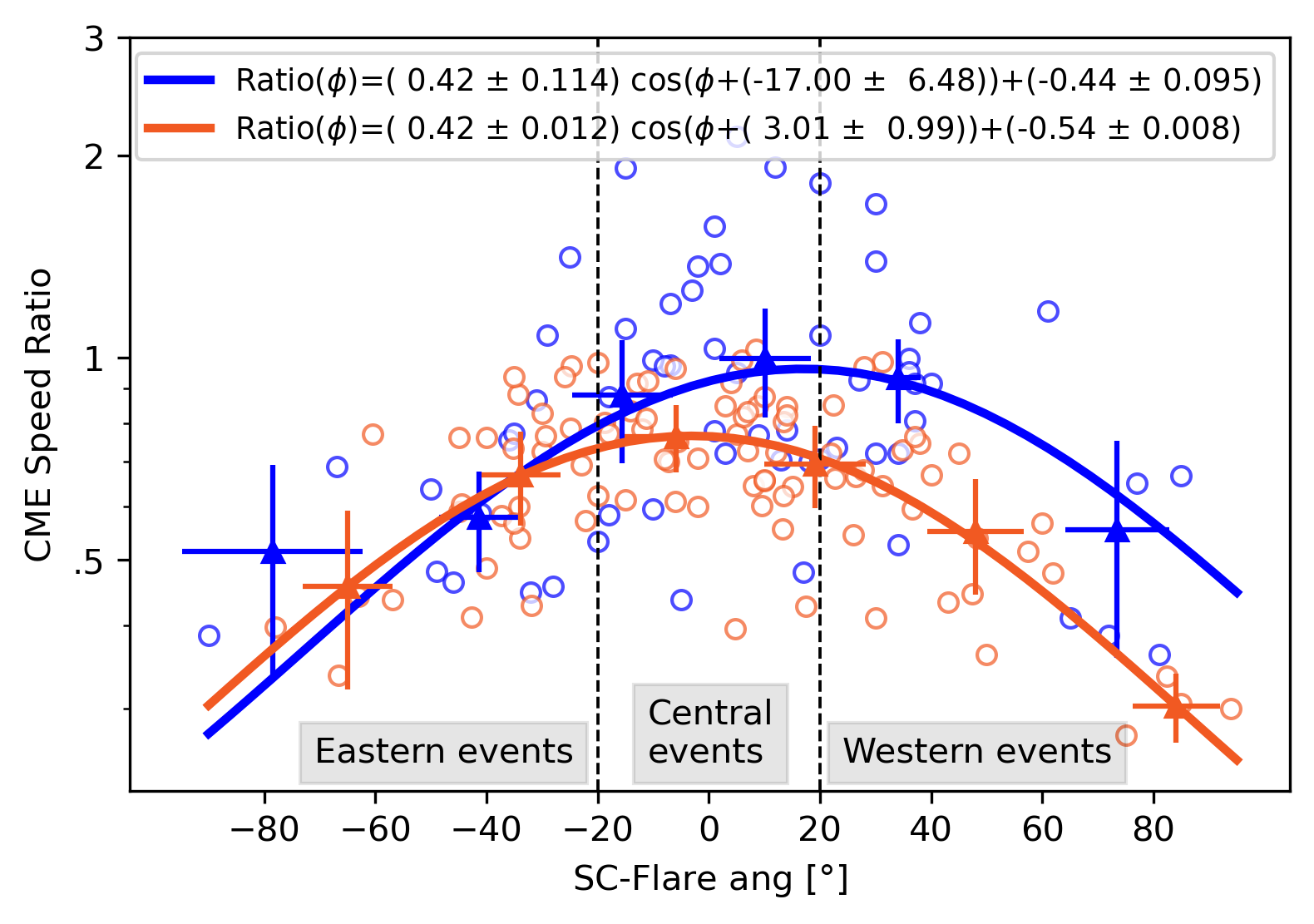} 
\centering
\caption{CME speed ratio (CME average transit speed/ CME near-Sun speed) versus spacecraft-flare angle of ESP events observed during SCs 23 (blue) and 24 (orange).}
\label{fig:Ratio}
\end{figure}

Figure \ref{fig:Ratio} displays the $CME\ Speed\ Ratio$ in relation to the angle between the spacecraft and the flare for ESP events observed during SCs 23 (blue) and 24 (orange). The plot compares the near-Sun speed of a CME at 21.5 $R_{Sun}$ to its average transit speed at different longitudes for each solar cycle. In SC 23, for example, at 15\degree, the CME average transit speed is, on average, $\sim$90\% of the estimated value at $\sim$21.5 $R_{Sun}$; whereas at -50\degree\ it is $\sim$60\%. The events of each SC were divided into six bins, and each bin's mean and standard errors are shown as triangles and crosses, respectively. The mean values of the bins were used to fit the following equation:
\begin{equation}
CME\ Speed\ Ratio(\phi)=V_0cos(\phi-\phi_0)+V_1 ,
\end{equation}

Here, $V_0$ represents the cosine amplitude of 0.42±0.114 and 0.42±0.012 for SC 23 and SC 24 events, respectively. $\phi$ represents the longitude between the observer and the CME solar source, while $\phi_0$ represents the offset longitude of the shock front nose and is -17.00±6.48\degree and 3.01±0.99\degree\ for SC 23 and SC 24 events respectively. Lastly, $V_1$ is an offset fitting constant of -0.44±0.095 and -0.54±0.008 for SC 23 and SC 24 events, respectively. Fits to individual data points yield similar results. SC 23 events generally have a higher speed ratio than SC 24 events along the longitudinal distribution.  However, it's important to note that the eastern events display a similar speed ratio longitude distribution during both SCs. Meanwhile, the western events show a significant difference, with SC 23 events having a higher speed ratio than SC 24 events. It is noteworthy that the CME speed ratio shown in Figure \ref{fig:Ratio} is correlated to the number of Eastern and Western ESP events displayed in Figure \ref{fig:hist}. During SC 23, Western events are more dominant than Eastern events, but during SC 24, the number of Eastern and Western events is similar. It is also important to note that some of the speed ratios of SC 23 CME are greater than 1. There could be various reasons, such as slow CMEs accelerated by a faster solar wind, plane-of-sky projection effects, or an error in the near-Sun CME speed estimate. Estimating these speeds during SC 23 was more challenging due to the lack of longitudinally separated observing s/c than during SC 24.

\begin{figure}[ht!]
\includegraphics[width=.6\linewidth]{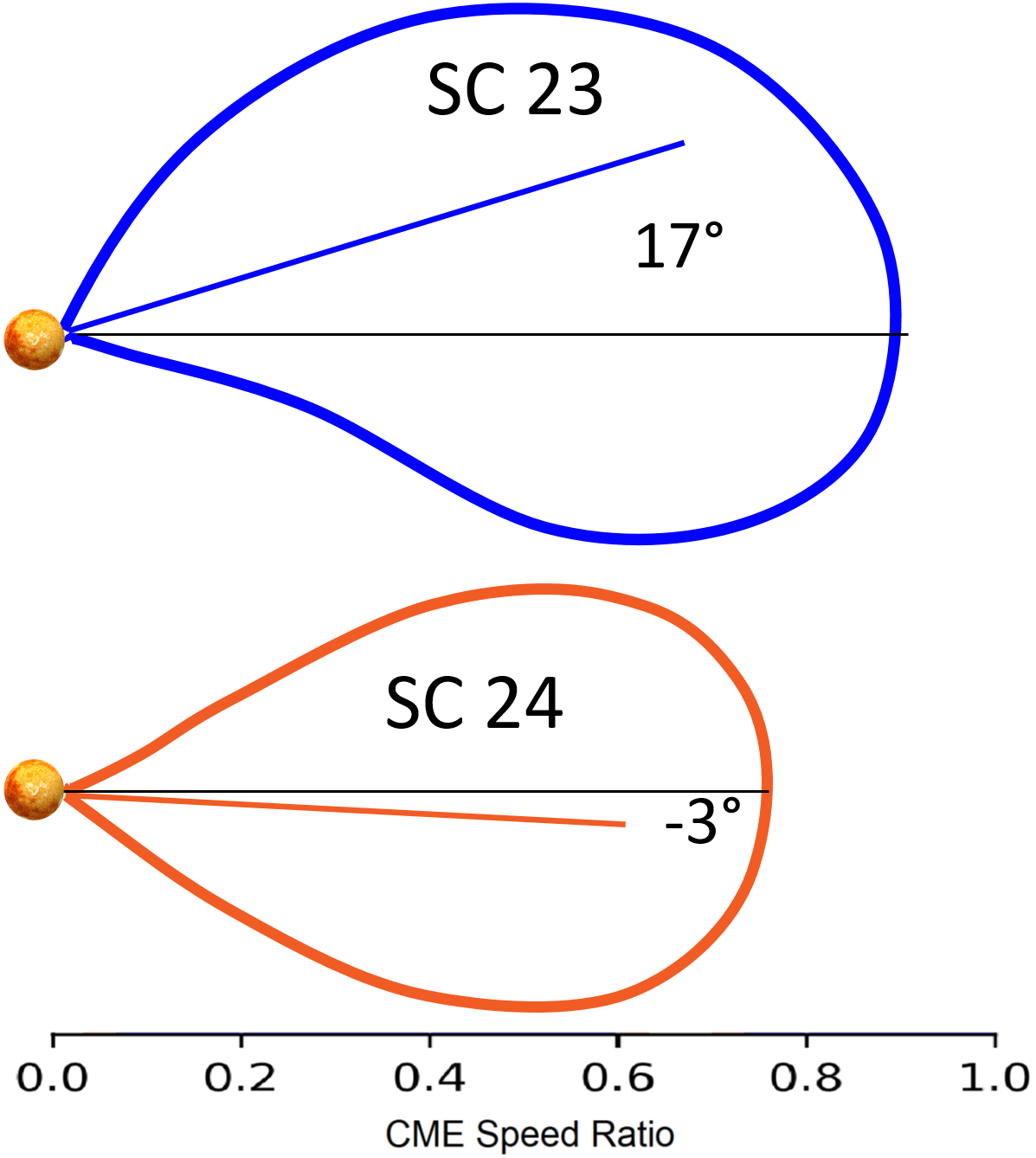} 
\centering
\caption{Average angular deflection and deceleration of the driving CMEs of ESP events observed during SCs 23 (blue) and 24 (orange).}
\label{fig:CMEAng}
\end{figure}

Our analysis shows a change in the deflection of CME events from approximately 17 degrees west during SC 23 to around 3 degrees east during SC 24, as depicted in Figures \ref{fig:Ratio} and \ref{fig:CMEAng}. We also observed a significant increase in the deceleration of SC 24 CME events from Figure \ref{fig:Ratio}. These findings suggest that physical mechanisms may be involved in these changes, which have been discussed in studies such as \citet{Wang2004} and \citet{Jian_2018}. Further investigation is necessary to understand the implications of these discoveries fully.

\begin{figure*}[hb]
\includegraphics[width=.9\textwidth]{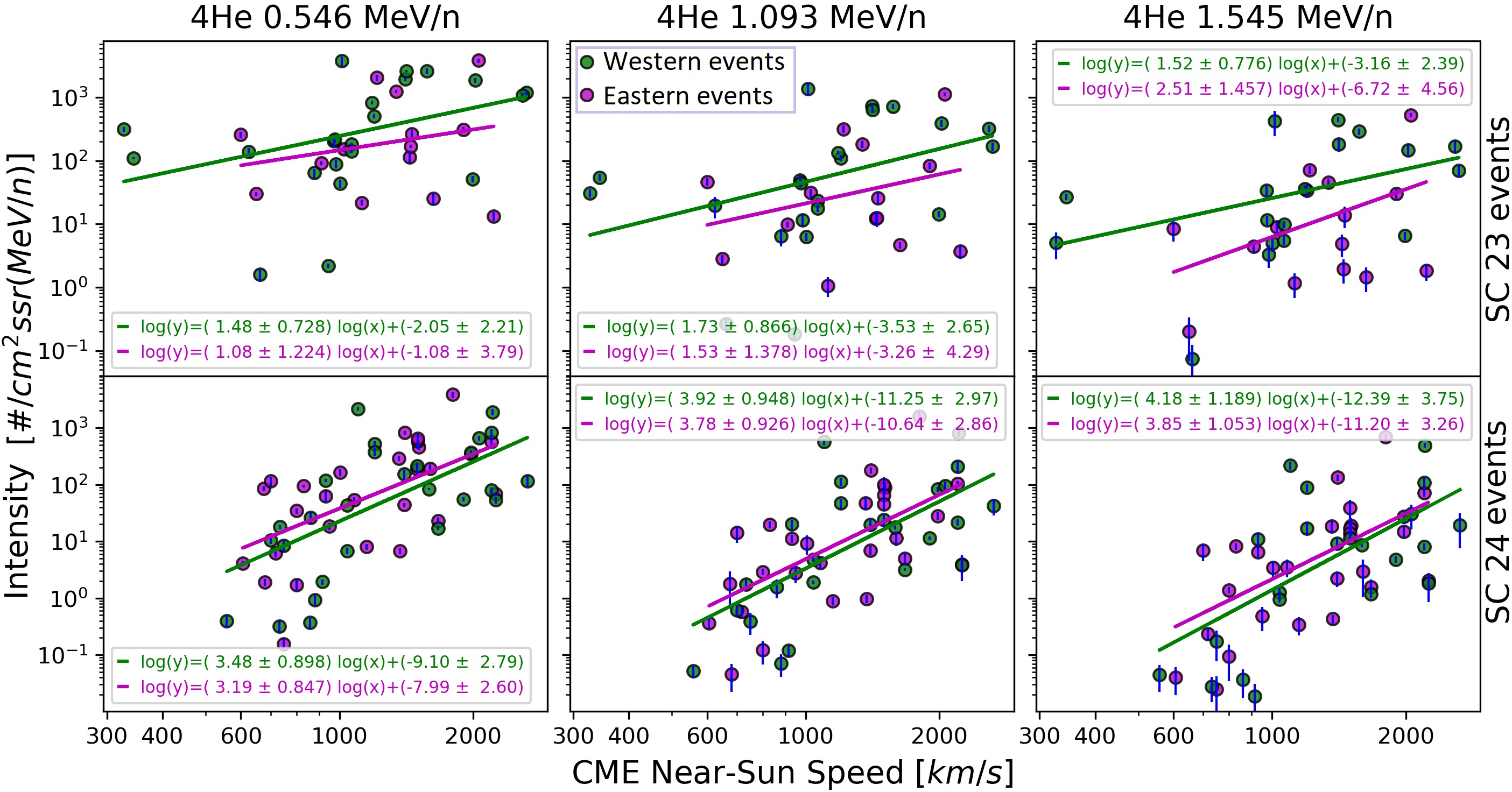} 
\centering
\caption{Peak intensity of He ion ESP observations versus CME near-Sun speed at $0.546$ MeV/n (left column), $1.093$ MeV/n (middle column) and $1.545$ MeV/n (right column) during SCs 23 (top panel) and 24 (bottom panel). The magenta dots and lines represent Eastern events and fit, respectively, while the green dots and lines represent the same for Western events.}
\label{fig:int_near}
\end{figure*}

\begin{figure*}[h]
\includegraphics[width=.9\linewidth]{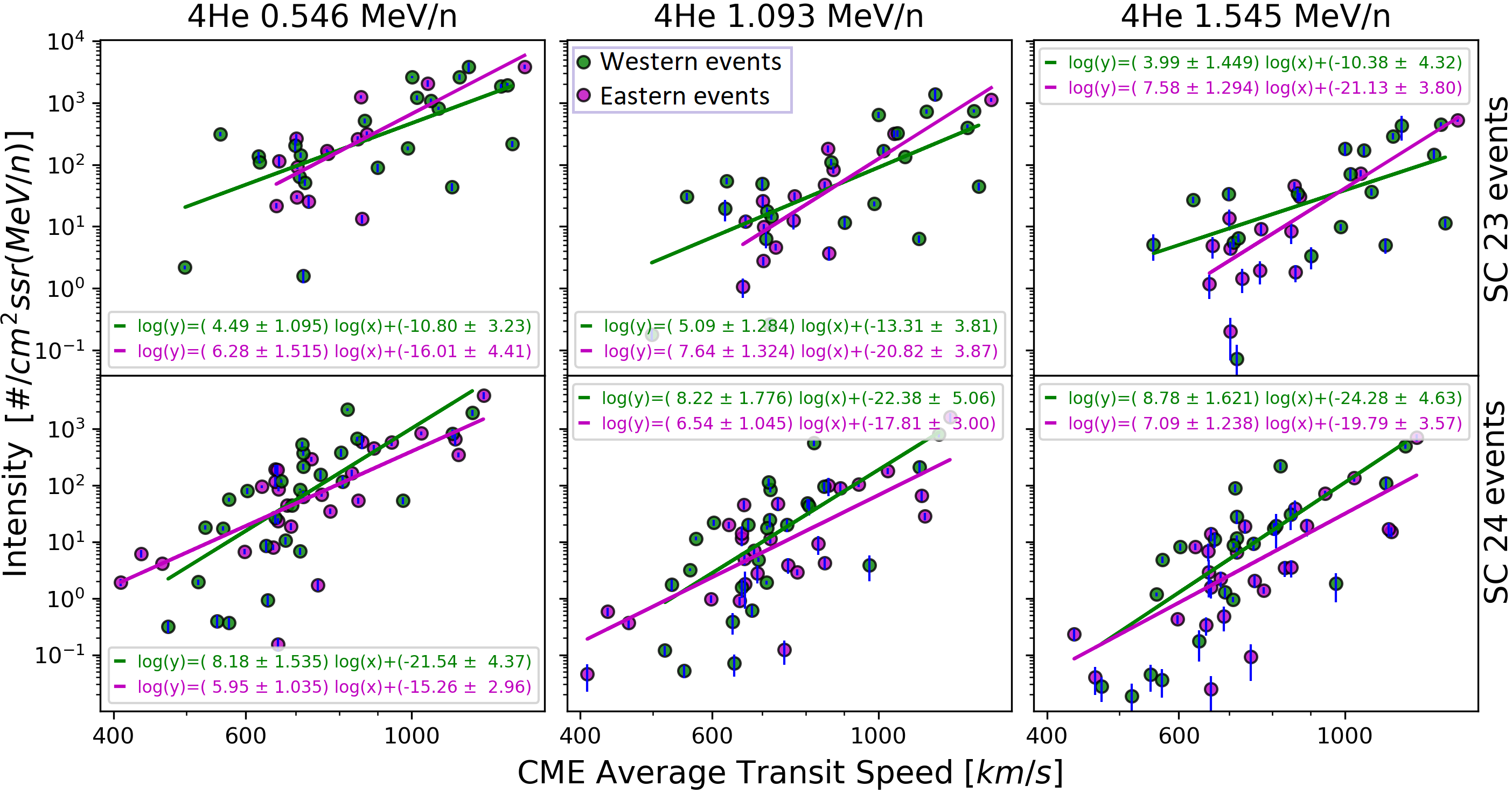} 
\centering
\caption{Peak intensity of He ion ESP observations versus CME average transit speed at $0.546$ MeV/n (left column), $1.093$ MeV/n (middle column) and $1.545$ MeV/n (right column) during SCs 23 (top panel) and 24 (bottom panel). The magenta dots and lines represent Eastern events and fit, respectively, while the green dots and lines represent the same for Western events.}
\label{fig:int_tran}
\end{figure*}

For each ESP event, we identified the intensity peak value between a $\pm 3$-hr window for each energy bin in the range from $\sim$ 0.1 - 3 MeV/nucleon for He, O, and Fe energetic ions. Figures \ref{fig:int_near} and \ref{fig:int_tran} show the peak intensity at $\sim 0.5$, $\sim 1.1$ and $\sim 1.5$ MeV nucleon$^{-1}$ of He-ion ESP events observed during SCs 23 (top panel) and 24 (bottom panel) versus the CME near-Sun and average transit speeds, respectively. Eastern events (in magenta) and Western events (in green) are shown with their fit lines.

During SC 23 ESP events, the peak intensities of western ESP events shown in Figure \ref{fig:int_near} are approximately five times higher than those in the east in all three energy bins. However, during SC 24, this feature is not observed; in fact, Eastern events generally have similar (slightly higher) peak intensity to Western events.

In Figure \ref{fig:int_tran}, it is noticeable that the fit lines of SC 23 Eastern events have a steeper slope compared to those of Western events. However, for SC 24 ESP events, a contradictory trend is observed. On the other hand, a study conducted by \citet{ding2023} reveals that when combining SC 23 and SC 24 events, Eastern ESP events exhibit slightly higher intensities than Western events based on their CME transit speed.

\begin{figure}[ht!]
\includegraphics[width=\linewidth]{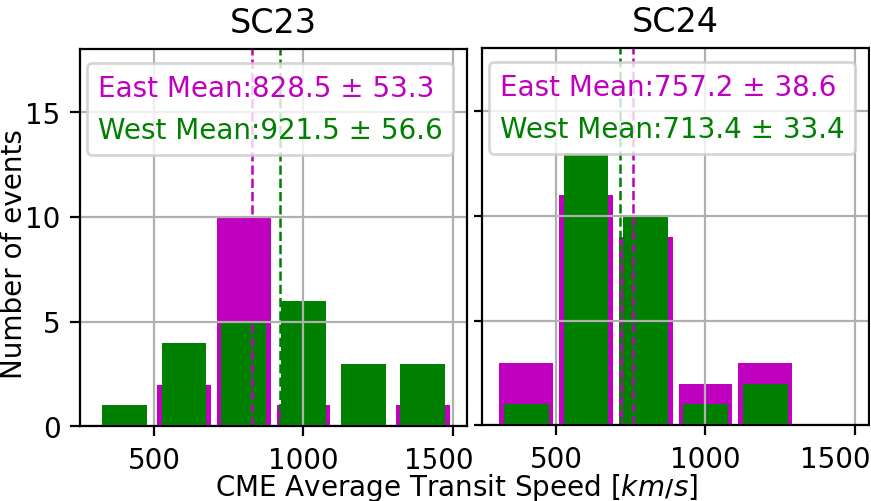} 
\centering
\caption{Distribution of CME average transit speeds in Eastern (purple) and Western (green) ESP events during SCs 23 (left) and 24 (right).}
\label{fig:EWAvg}
\end{figure}


\begin{figure*}
\includegraphics[width=1\linewidth]{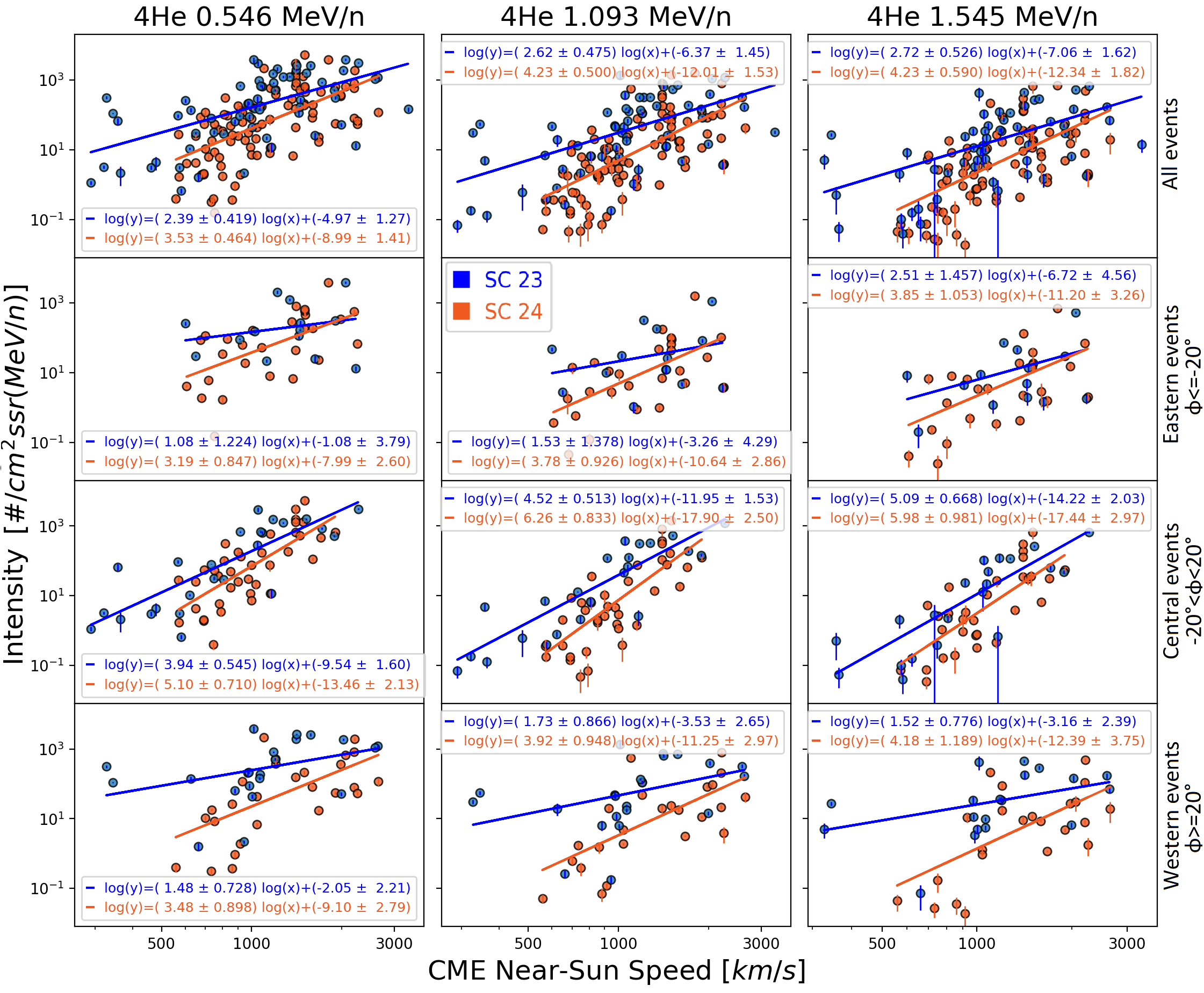} 
\centering
\caption{Peak intensity of He ion ESP observations versus CME near-Sun speed at $0.546$ MeV/n (left column), $1.093$ MeV/n (middle column) and $1.545$ MeV/n (right column) during SCs 23 and 24. All events of both SCs are shown in the first row. Eastern ($\phi<=-20\degree$), central  ($-20\degree<\phi<20\degree$), and Western ($\phi>=20\degree$) events are shown in rows 2, 3, and 4, respectively. Blue (orange) dots and lines denote SC 23 (SC 24) observations and fits, respectively.}
\label{fig:NSCME}
\end{figure*}
\begin{figure*}
\includegraphics[width=1\linewidth]{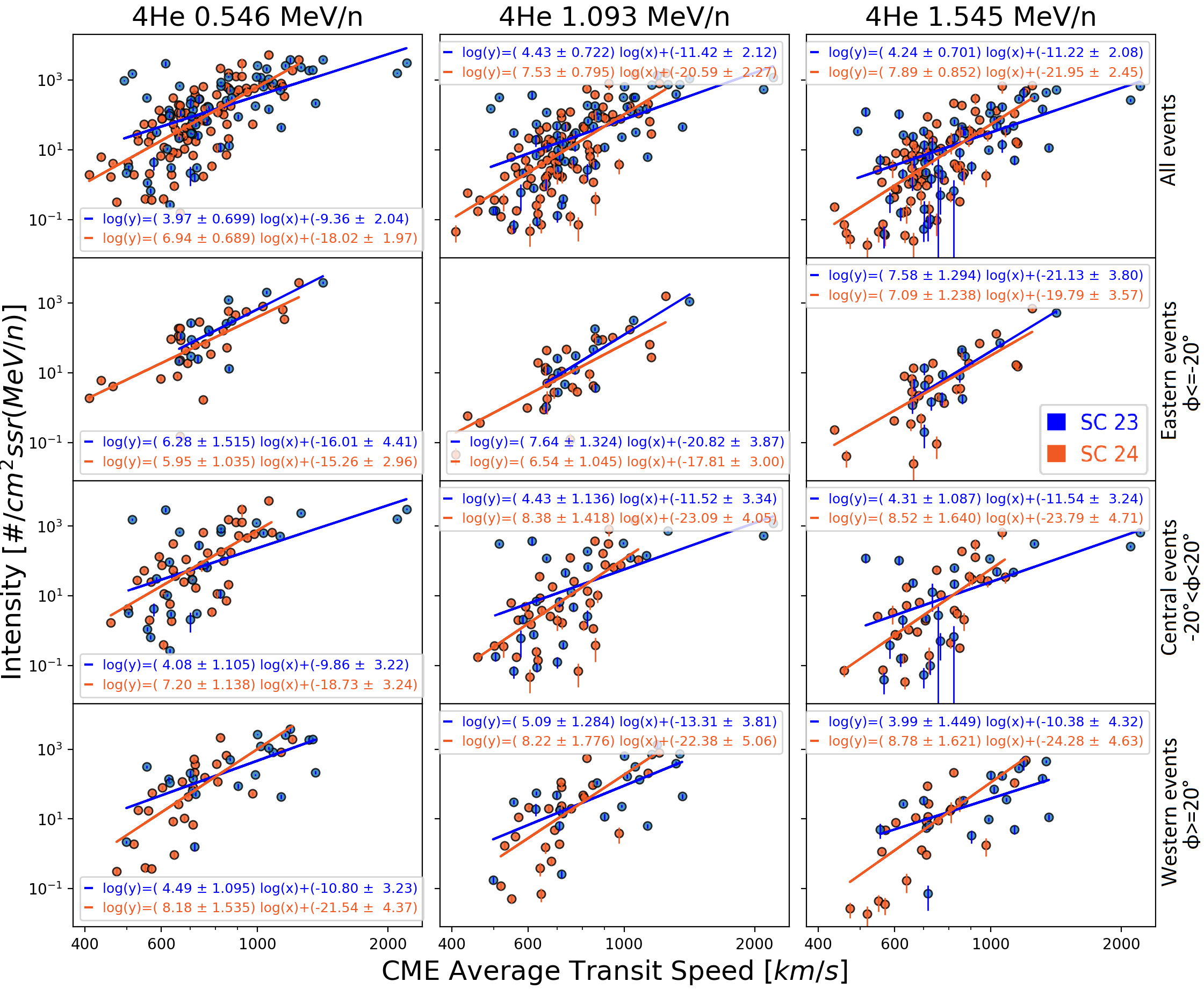} 
\centering
\caption{Peak intensity of He ion ESP observations versus CME average transit speed at $0.546$ MeV/n (left column), $1.093$ MeV/n (middle column) and $1.545$ MeV/n (right column) during SCs 23 and 24. All events of both SCs are displayed in the first row. Eastern ($\phi<=-20\degree$), central  ($-20\degree<\phi<20\degree$), and Western ($\phi>=20\degree$) events are shown in rows 2, 3, and 4, respectively. The blue dots and lines represent the SC 23 observations and fits, respectively, while the orange dots and lines represent the same for SC 24.}
\label{fig:tran}
\end{figure*}

Figure \ref{fig:int_near} indicates a reversal in the east-west asymmetry of the ESP peak intensity versus the near-Sun CME speed distribution. Additionally, Figure \ref{fig:int_tran} suggests that the trend of the slopes of the peak intensity versus the CME average transit speed has also reversed. Thus, in SC 23, the intensity of Western events was higher in Figure \ref{fig:int_near}, and they had faster CME transit speeds compared to Eastern events in Figures \ref{fig:Ratio} and \ref{fig:EWAvg}. However, the scenario was reversed in SC 24, where Eastern events had higher intensities and faster CME transit speeds than Western events.

Figures \ref{fig:NSCME} and \ref{fig:tran} compare the peak intensities of each heliolongitude group (Eastern, Central, and Western) during both SCs with respect to their CME near-Sun and average transit speeds, respectively. The figures help illustrate the variations in east-west peak intensity and their respective relationship with the CME near-Sun and average transit speeds.

The plot in Figure \ref{fig:NSCME} illustrates the peak intensity of He-ion ESP events at around 0.5, 1.1, and 1.5 MeV nucleon$^{-1}$ for SCs 23 (blue) and 24 (orange) in relation to the CME speed near the Sun. The top row displays all the events for both SCs, while the second, third, and fourth rows show eastern, central, and western events, respectively. In general, the fits (blue and orange lines) for all events show higher intensities in SC 23 events than for SC 24. The same is true across all three energy ranges of He ions, similar to O and Fe (not shown).

Figure \ref{fig:tran} shows the peak intensity at $\sim 0.5$, $\sim 1.1$ and $\sim 1.5$ MeV nucleon$^{-1}$ of He-ion ESP events observed during SCs 23 (blue) and 24 (orange) versus the CME average transit speed. The top row shows that all the events' fit (blue and orange lines) are steeper for SC 24 than for SC 23 events, as reported by \citet{SantaFe2023}. However, the slope and y-intercept values of Eastern events in both SCs are more similar than those of Central and Western events. Western events in SC 24 have steeper fit lines than SC 23.

We point out that a direct correlation exists between the speed ratio of CMEs for the Eastern, Central, and Western clusters of ESP events, their respective peak intensities, and the CME near-Sun and average transit speeds. The Eastern ESP events of both SCs display a degree of similarity in various aspects. Figure \ref{fig:Ratio} indicates that the Eastern CME speed ratio heliolongitude distributions are similar in both SCs. Similarly, Figures \ref{fig:NSCME} and \ref{fig:tran} demonstrate that the Eastern peak intensities exhibit comparable patterns for high-speed ESP events. The analysis of the Western ESP events reveals that SC 23 events exhibit a higher CME speed ratio than SC 24 events, as demonstrated in Figure \ref{fig:Ratio}. Similarly, the peak intensities of SC 23 events are higher than those of SC 24, as illustrated in Figures \ref{fig:NSCME}. Furthermore, Figures \ref{fig:tran} indicate that SC 23 events demonstrate higher CME average transit speeds than SC 24 events. In other words, the steeper slope of the Western ESP events of SC 24 shown in Figure \ref{fig:tran} can be explained by their stronger deceleration compared to SC 23 events as shown in Figure \ref{fig:Ratio}.

\section{Discussion}

ESP events profoundly impact space weather and pose a significant threat to space-based technologies \citep{Barth2003}. As such, a comprehensive investigation is paramount to fully comprehend the underlying mechanisms that drive these events \citep{Cucinotta2010}. It has been suggested that the differences in peak intensities of ESP events between the east and west could be influenced by the characteristics of the IP CME shock and the ambient conditions \citep{Sarris1985}. \citet{ding2023} shows the peak intensity at the eastern flank of CME shock is generally larger than the one at the western flank. However, our results reveal a reversal of the east-west asymmetry in the intensity of ESP peak events from SC 23 to SC 24. It highlights the need for further research to understand better the complex dynamics of ESP events and their relationship to solar activity.

Our analysis found that the distribution of CME speed ratio with respect to observing s/c heliolongitude (Figure \ref{fig:Ratio}) plays a crucial role in the east-west asymmetry of the intensities of ESP events. Based on the distribution, it is observed that the average CME during SC 23 would deflect from the western heliosphere with an angle of $\sim17\pm6\degree$ towards Earth. On the other hand, for SC 24, the CME would deflect from the eastern heliosphere with an angle of $\sim3\pm1\degree$. Additionally, during SC 23, on average, the nose shock-front of a CME had a transit speed of roughly $90\%$ of its initial speed near the Sun, and experienced weaker deceleration on both flanks. However, for SC 24, the nose shock-front of the average CME had an average transit speed of approximately $75\%$ of its initial speed near the Sun, experiencing stronger deceleration on both flanks. These observations support the hypothesis that weak CMEs are deflected to the west and fast CMEs to the east by the Parker spiral, as proposed by \citet{Wang2004}, and that SC 24 energetic particle events were weaker than those of SC 23, as reported by \citet{Jian_2018}.

The variations in the speed evolution and deflection of CME events during SCs 23 and 24 have led to notable differences in the speed ratios of Eastern and Western events, as shown in Figure \ref{fig:Ratio}. Specifically, it has been observed that the Eastern events had comparable speed ratios in both cycles. However, the Western events of SC 23 exhibited higher speed ratios than those of SC 24. As a result, the peak intensities distributions in relation to the CME average transit speed are similar for Eastern ESP events during both SCs, as shown in Figure \ref{fig:tran}. Meanwhile, for Western events, SC 23 events had higher peak intensities in relation to the CME near-Sun speed than SC 24, as shown in Figure \ref{fig:NSCME}.. In addition, the more substantial deceleration of Western SC 24 events depicted in Figure \ref{fig:Ratio} is also represented in Figure \ref{fig:tran} by the steeper slopes of their fitting lines compared to Western SC 23 events.

The results indicate that the primary factor responsible for the reversal of the east-west asymmetry in the intensity of ESP events between SC 23 and 24 is the shift in the peak of the CME speed ratio heliolongitude distribution from $\sim17\pm6\degree$ to $\sim-3\pm1\degree$, as illustrated in Figures \ref{fig:Ratio} and \ref{fig:CMEAng}. This finding suggests that the shift in the peak of the CME speed ratio is a crucial determinant of the reversal of the asymmetry.
It is essential to conduct further research to determine whether the variable behavior of CME events during different solar cycles is influenced by the reversal of magnetic fields \citep{Gopalswamy_2003}, the strength of the solar cycle \citep{SantaFe2023}, the strength of the interplanetary magnetic field \citep{Jian_2018}, the CME interchange reconnection \citep{Winslow_2023} or other underlying mechanisms. This is especially important due to the significant impact that CMEs have on space weather, particularly on space assets and astronauts.

In conclusion, our study has found that there has been a reversal of the east-west asymmetry in the intensity of ESP events from SC 23 to SC 24. We have also discovered that the primary reason for this asymmetry reversal is the  heliolongitude shift of the peak of the distribution of the CME speed ratio from west to east. However, more research is necessary to understand the mechanisms underlying this heliolongitude shift.

\section*{Acknowledgements}

Work at SwRI was supported by NASA grants NNX17AI17G, 80NSSC23K0975 and 80NSSC23K1470. LKJ was supported by NASA's STEREO mission and HGI Grant 80NSSC23K044. We would like to express our gratitude to the teams behind STEREO/SIT, PLASTIC, MAG, ACE/ULEIS, SIS, SWEPAM, and MAG for generously providing their data for public use. We thank NASA's Community Coordinated Modeling Center for making the WSA-ENLIL simulations publicly available through the Space Weather Database of Notifications, Knowledge, Information (DONKI).

\bibliography{sample63}{}
\bibliographystyle{aasjournal}

\end{document}